\documentclass[aps,pra,a4paper,twocolumn,notitlepage]{revtex4-1}

\usepackage{amsmath}
\usepackage{bm}
\usepackage{graphicx}

\begin{document}

\title{Optical properties of spherulite opals}
\author{Venkata Jayasurya Yallapragada}
\email{venkata-jayasurya.yallapragada@weizmann.ac.il}
\author{Dan Oron}
\affiliation{Department of Physics of Complex Systems, Weizmann Institute of Science, 234 Herzl Street, Rehovot 7610001, Israel}

\begin{abstract}
Spherulites are birefringent sturctures with spherical symmetry, which are typically observed in crystallized polymers. We compute the band structure of opals made of close-packed assemblies of highly birefringent spherulites. We demonstrate that spherulitic birefringence of constituent spheres does not affect the symmetries of an opal, and yet significantly affects the dispersion of eigenmodes, leading to new pseudogaps in sections of the band structure, and consequently enhanced reflectivity.
\end{abstract}
\maketitle

Photonic crystals \cite{yab1987,book_joannopoulos,busch-john}, which are structures with a periodically varying refractive index, have been among the most studied systems in optics during the last few decades. This periodicity in refractive index results in the formation of photonic band gaps and modes with tailorable dispersion properties. The ability to engineer the propagation of light through a photonic crystal has resulted in a number of applications in the design of optical cavities, waveguides and optical fibers. When the band gap occurs for a certain frequency range for propagation along all directions in the photonic crystals, a complete band gap is said to be present. Complete photonic bandgaps require a very large refractive index contrast which is typically not possible for visible light using structures made of commonly available transparent dielectrics. However, partial band gaps, occurring only for propagation of light along certain directions in the crystal, can be obtained. 

Opals are naturally occurring photonic crystals, and consist of a close packed assemblies of dielectric nanospheres. Several synthetic analogs, made using colloidal self assembly techniques, have also been demonstrated \cite{xia_review, kim2011self}. The optical properties of a synthetic opal can be engineered by changing the size, material properties and internal structure of the constituent nanospheres \cite{stein_com}. Opals also serve as precursors to realizing media with more interesting photonic band structures, such as inverse opals \cite{stein_adfm}. 

Optical anisotropy has been shown to facilitate complete band gaps in two dimensional photonic crystals \cite{Li1998_2d}, and significantly alter the band structure of three-dimensional arrays of nanospheres \cite{Li1998_3d, zabel_stroud}. Infiltration of photonic crystals with birefringent liquid crystals has been employed to demonstrate tunability of photonic band structure \cite{busch-john-lc}. Calculations of bandstructure of opals made of birefringent uniaxial spheres have been presented by Zabel and Stroud \cite{zabel_stroud}. In structures described by them, the optic axis of each sphere is aligned in the same direction. The difference in the effective index for different polarizations results in the splitting of the lowest frequency photonic bands which are degenerate in an FCC opal made of isotropic spheres. This can also be understood as a lifting of the degeneracy of these bands due to symmetry. A disadvantage of these birefringent sphere opals is that they cannot be realized using colloidal self-assembly, since they require precise orientation of the optic axis of each constituent sphere. 

\begin{figure}[t]
	\centering\includegraphics[width=3.4in]{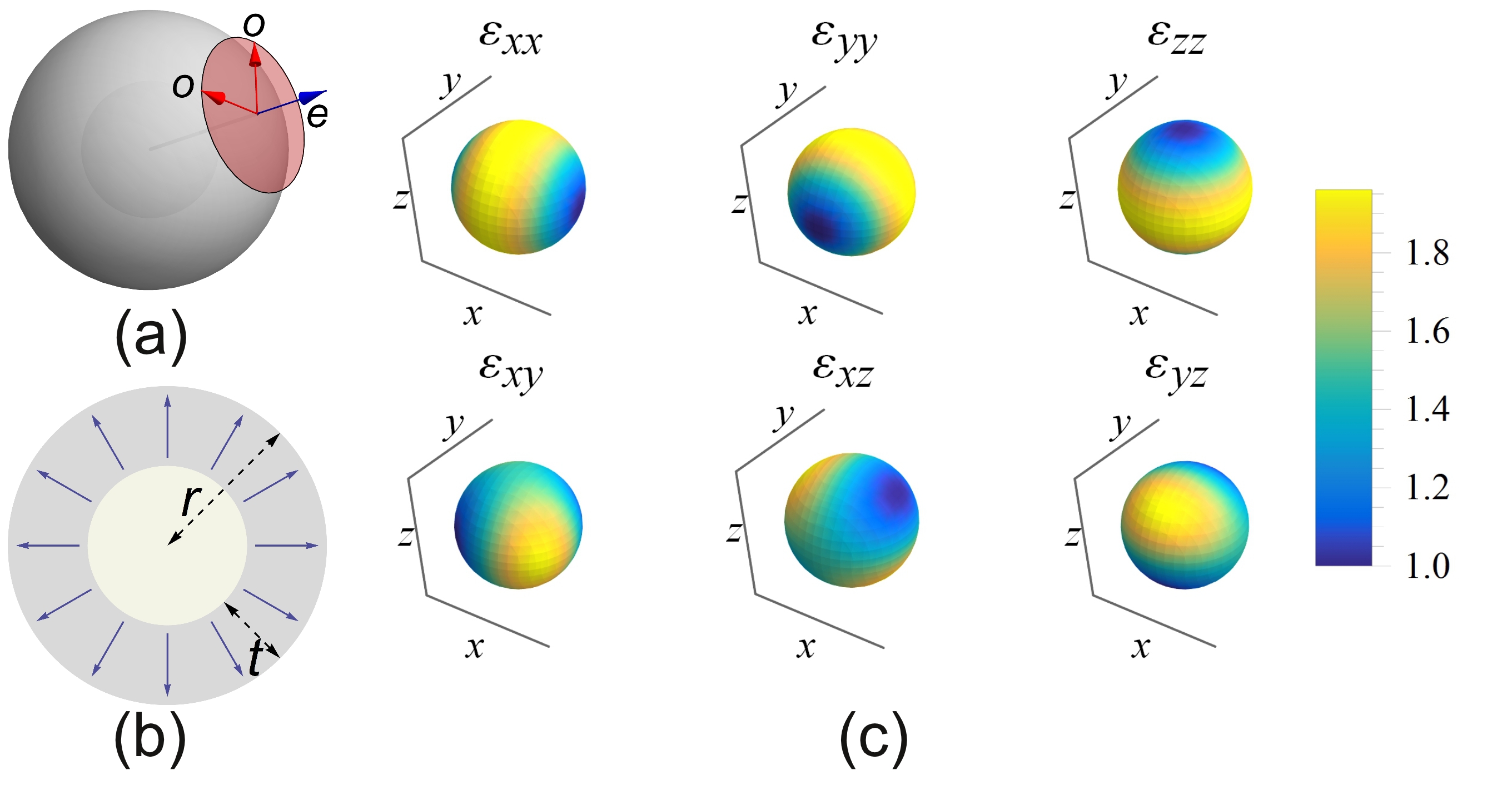}
	\caption{(a) Schematic of a spherulite showing the orientation of the optic axis (blue). o and e indicate ordinary (tangential) and extraordinary (radial) indices respectively (b) Cross section schematic of a core-shell spherulite of radius $r$ and shell thickness $t$. Solid blue arrows indicate the orientation of the optic axis within the shell. (c) The Cartesian components of the dielectric tensor inside the material of a spherulite with radial and tangential refractive indices $n_{r} = 1.0$ and $n_{t}=1.4$, respectively. 
 	}
	\label{fig_spherulites}
\end{figure}

Recently, opal-like assemblies of highly birefringent core-shell nanospheres have been reported in the eyes of decapod crustaceans \cite{ben_pnas}. These spheres have been found to consist of deep-subwavelength thickness platelets arranged in a lamellar fashion around a low index core. These platelets are made of crystalline isoxanthopterin (calculated refractive indices $n_{1} = 1. 89$, $n_{2}=2.02$, $n_{3}= 1.40$) and are oriented in such a way that the low index ($n_{3}=1.40$) axis always points radially outwards and the remaining axes are not oriented in a specific direction, effectively rendering the shell uniaxially birefringent \cite{decapod_paper}. Using Mie theory and FDTD calculations, the authors have shown that the spherulitic core-shell spheres outperform isotropic shells with similar refractive index as reflectors, exhibiting a high reflectivity over a wider band of wavelengths. 

These particles belong to a class of spherical crystalline objects known as spherulites, which are spheres made of uniaxial birefringent material with the optic axis oriented in the radial direction \cite{bower_book}, as shown in Fig.\ref{fig_spherulites}(a) and Fig.\ref{fig_spherulites}(b). This orientation of the optic axis ensures the spherulite possesses the same rotational symmetry as an isotropic sphere. The dielectric tensor of a spherulite in spherical coordinates has two unique elements, $\varepsilon_{rr} = n_{r}^2$ and $\varepsilon_{\theta\theta}= \varepsilon_{\phi\phi}= n_{t}^2$, where $n_{r}$ and $n_{t}$ are refractive indices for light polarized radial and tangential to the surface of the spherulite, respectively. The variation in Cartesian coordinates is more complex, and is shown in Fig.\ref{fig_spherulites}(c). The optical scattering properties of individual spherulites have been studied earlier, and results indicate that spherulites are capable of scattering more efficiently in certain directions than isotropic particles of similar index contrast \cite{roth_dignam, ni_spherulite, kiselev_tmatrix}. In the present article, we  explore the properties of close-packed assemblies of spherulites, which we hereafter refer to as spherulite-opals. We compare the photonic band structures and reflectivity spectra of spherulite opals with those of opals made of isotropic spheres, and also study the effects of optical anisotropy and the presence of a low index core on these structures.  

In order to study the effects of spherulitic birefringence on the optical properties of a spherulite opal, we have computed its photonic band structure, and compared it with opals made of isotropic spheres. In order to compute the band structure, we use the following eigenvalue equation, which is derived from Maxwell's equations \cite{book_joannopoulos}.
\begin{equation}
\nabla \times \left( \bm{\varepsilon}^{-1}(\mathbf{r})\nabla \times \mathbf{H} \right) = \frac{\omega^{2}}{c^{2}}\mathbf{H}
\label{eqn_H}
\end{equation}
Here, $\bm{\varepsilon}^{-1}(\mathbf{r})$ is the inverse dielectric tensor at each spatial coordinate; $\omega$ and $\mathbf{H}$ are the frequency (eigenvalue) and magnetic field (eigenvector), respectively, corresponding to the eigenmode. Equation \ref{eqn_H} is solved with periodic boundary conditions characteristic of the FCC lattice.

\begin{figure}[t]
	\centering\includegraphics[width=3.4in]{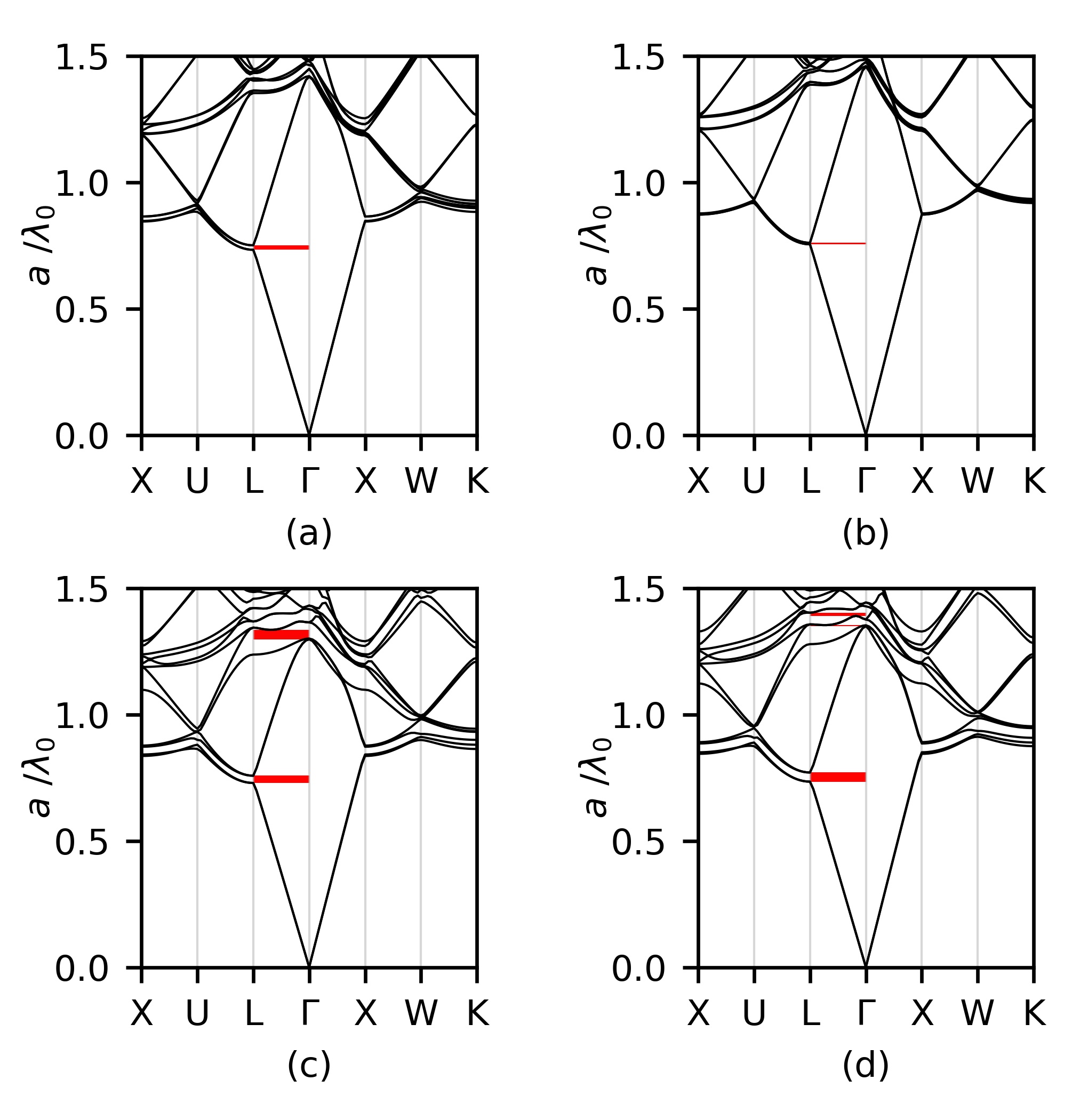}
	\caption{
		Photonic band structure of an opal consisting of a close packed FCC arrangement of 
		(a) dielectric spheres of refractive index $n = 1.267$. 
		(b) core-shell dielectric spheres, with a core index of $n_{c}=1$. 
		(c) spherulitic spheres of $n_t=1.4$ and $n_r = 1.0$, and 
		(d) spherulitic core-shell spheres of $n_t=1.4$ and $n_r = 1.0$ and core index $n_c = 1.0$.
		The red bands indicate the pseudogaps in the $\Gamma$L section of the Brillouin zone. 
	}
	\label{fig_bands}
\end{figure}

For the results that follow, we computed the eigenmodes of Maxwell's equations with periodic boundary conditions in a plane wave basis using the freely available software package MPB via its Python interface \cite{mpb}. Our choice of material parameters is motivated by the index contrast between isoxanthopterin and water. We use ordinary and extraordinary indices $n_o = 1.4$ and $n_e = 1.0$ for the uniaxial birefringent material that constitutes the spherulites. For comparison we use opals made of optically isotropic spheres of refractive index $n = 1.267$, obtained by averaging the three indices of the birefringent material

We first calculate the well-known bandstructure of an opal made of isotropic dielectric spheres of radius $r$ and refractive index $n=1.267$ arranged in a close packed FCC lattice, which is shown in Fig. \ref{fig_bands}(a). Frequencies in the band structure are expressed in units of $a/\lambda_{0}$, where $a$ is the length of the edge of the FCC cubic unit cell and $\lambda_{0}$ is the free space wavelength. When such opals are formed by self assembly from a colloid, interfaces parallel to the (111) plane are most commonly expressed. Features in the reflectivity spectrum at such interfaces are affected primarily by the band structure in $\Gamma$L direction \cite{bogomolov, pavarini}.  Pseudogaps in the $\Gamma$L section of the Brillouin zone, marked in the band structures in Fig. \ref{fig_bands}, manifest themselves as peaks in reflectivity. The introduction of the core of radius $r/2$ and refractive index $n=1$ results in a reduction of the effective index of the photonic crystal, and we therefore observe an overall increase in the mode frequencies. In addition, the widths of the pseudogaps are reduced, as shown in Fig. \ref{fig_bands}(b).

The band structure of a spherulite opal with $n_{r}=1.0$ and $n_{t}=1.4$, computed with the same technique as the isotropic sphere opal, is shown in Fig. \ref{fig_bands}(c). The $\Gamma$L pseudogaps are more pronounced and there is a pseudogap in the high frequency regime between bands 4 and 6. Also, the modes around the pseudogap are noticeably flatter than in case of the isotropic opal, indicating low group velocities. The introduction of an $n=1$ core of radius $r/2$ in the spherulite opal reduces the width of this gap, but another gap is introduced between bands 7 and 8, as illustrated in Fig. \ref{fig_bands}(d). These pseudogaps occur for wavelengths equal to the lattice spacing, and result from the refractive index contrast within each spherulite.

Since a spherulite has the same rotational symmetry as a sphere, the symmetry of the primitive FCC unit cell is unaffected. The dielectric tensor $\bm{\varepsilon}(\mathbf{r})$ within an isotropic dielectric sphere in spherical coordinates $(r,\theta,\phi)$ is diagonal, with $\varepsilon_{rr}=\varepsilon_{\theta\theta}=\varepsilon_{\phi\phi}=n^2$. Any modification to the dielectric tensor of an isotropic medium that retains its diagonal nature and ensures $\varepsilon_{\theta\theta}= \varepsilon_{\phi\phi}$ results in a spherulite and does not affect the degeneracy of the $\Gamma$L modes. In addition, the symmetry criteria that determine the ability of a mode to couple to incident light remain unchanged \cite{sakoda2004optical, sakoda_prb}. However, there is increased contrast of refractive index within the sphere felt by electric fields polarized along the tangential and radial direction in each spherulite. Therefore the reason for the formation and enhancement of pseudogaps is qualitatively different from the case of an opal constructed from oriented birefringent spheres \cite{zabel_stroud}, in which mode symmetries and degeneracies are significantly altered due to anisotropy. 

\begin{figure}[t]
	\centering\includegraphics[width=3.4in]{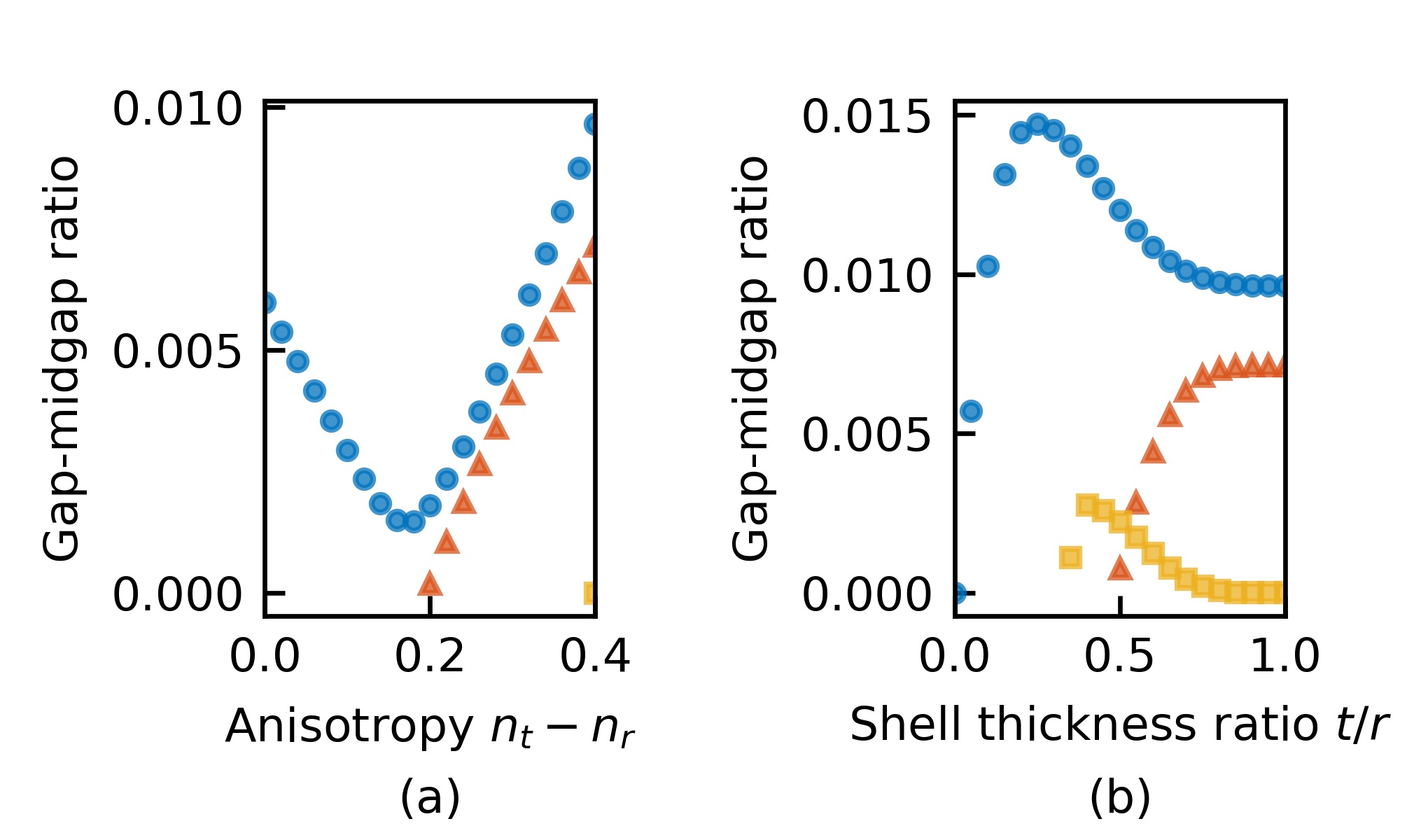}
	\caption{
		(a) The dependence of the gap-midgap ratio of the pseudogaps in the $\Gamma$L section of the Brillouin zone for a solid sphere opal as anisotropy is increased. As the anisotropy is increased, higher order gaps begin to open up.
		(b) A plot similar to (a) showing the variation of the gap-midgap ratio of various pseudogaps with shell thickness ($t$) in a core shell spherulite opal. Thickness $t=0$ corresponds to free space. }
	\label{fig_gaps}

\end{figure}

We first calculated the dependence of the width of these gaps on the magnitude of spherulitic anisotropy of each sphere. The lowest order pseudogap first decreases in width and then increases with increasing anisotropy. This is shown in Fig. \ref{fig_gaps}(a), where the ratio of the pseudogap width to the frequency at the center of the gap (the gap-midgap ratio) is plotted. An intuitive description of this variation is as follows \cite{book_joannopoulos}. The lowest band possesses a field distribution that concentrates the fields in regions of high dielectric function i.e. the dielectric spheres. The next lowest mode is orthogonal to it, therefore has higher concentration of its field in the low-index voids between spheres. At the edges of the Brillouin zone (for example at points X and L), this leads to a difference in the frequency of these bands. The presence of a low index region within the sphere, due to the lower index for light polarized in the radial direction, reduces the frequency difference, since it evens out the effect of the voids. Therefore the lowest pseudogap reduces in width as anisotropy is increased. However, at very large values of anisotropy, this effect is overwhelmed by the larger index contrast between radial and tangential refractive indices, which leads to increased pseudogap.

The bands that lie between $a/\lambda_{0}=1.2$ and $a/\lambda_{0}=1.4$ at $\Gamma$ involve diffraction from the (200) and (111) planes of the opal. The spacing of frequencies of these bands is influenced by periodicities twice the spatial frequency of the lowest pseudogap, and it is therefore intuitive that the intrinsic variation of refractive index across the spherulite leads to increased separation of mode frequencies at the zone edge. This intuition is supported by the similarity of frequencies of onset of the 4-6 pseudogap and the minimum of the 2-3 pseudogap width. The dependence of the pseudogap widths on the ratio of shell thickness to the radius is plotted in Fig.\ref{fig_gaps}(b). At $t/r \cong 0.6$, both 4-6 and 7-8 gaps exist. For shells thicker than about $t = 0.75r$, the gap widths are quite robust to changes in the shell thickness. In a spherulite with $n_t = 1.4$ and $n_r=1.0$, the dominant periodicity is from the contrast between the radial and tangential refractive indices.  As a consequence the inclusion of a low index ($n = 1.0$) core in the spherulite does not affect the pseudogaps in the band structure as much as it does in the case of an isotropic sphere.

\begin{figure}[t]
	\centering\includegraphics[width=3.4in]{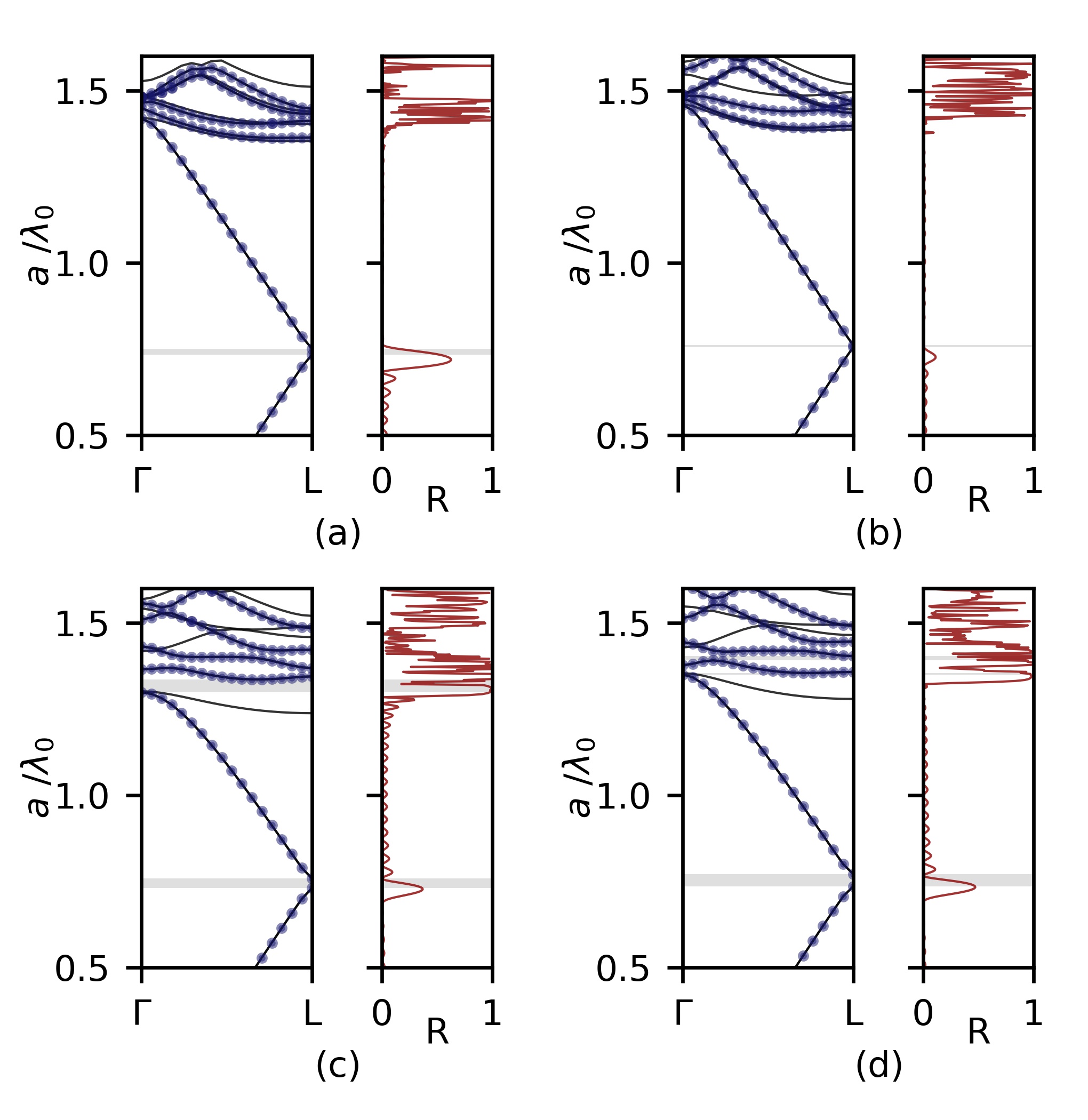}
	\caption{
		Bands in the $\Gamma$L direction and normal-incidence reflectivity as a function of frequency for opals made of 
		(a) solid isotropic dielectric spheres of refractive index $n = 1.267$, 
		(b) Core-shell isotropic spheres of shell index $n = 1.267$ and core index $n_{c}=1.0$, 
		(c) Solid spherulites with refractive indices $n_t = 1.4$ and $n_r = 1.0$, and 
		(d) core shell spherulites with shell index $n_t = 1.4$ and $n_r = 1.0$ and core index $n_c = 1.0$.
		In the band structure plots, the circular markers indicate the bands that the incident light at normal indicence can couple to. The gray areas indicate the pseudogaps shown in Fig. \ref{fig_bands}.
	}
	\label{fig_ref}
\end{figure}

In order to visualize the effects of the modified band structure of spherulite opals on reflection properties, we computed the reflectivity of an interface parallel to the (111) plane of the opal at normal incidence using the Finite Difference Time Domain (FDTD) technique \cite{taflove}. We used a commercial FDTD solver capable of handling inhomogeneous anisotropic media to perform these calculations \cite{lumerical}. The structure consisted of 18 layers of spheres, which is sufficient to fully express the features in reflectivity at frequencies above the fourth band. 

The reflectivity a (111) face of an isotropic sphere opal at normal incidence is shown in Fig. \ref{fig_ref}(a), together with the bands in the $\Gamma$L direction. The frequency range immediately above the fourth band contains several modes that couple to the incident light and the density of states is non-zero \cite{pavarini,agio}. The isolated peaks in reflectivity in this frequency range do not arise from pseudogaps, but result from band crossings and group velocity anomalies \cite{galisteo_high_energy}. Introduction of a low-index core of radius $r/2$ does not affect the qualitative features of the reflectivity spectrum, and further reduces the lowest $\Gamma$L pseudogap, as shown in Fig. \ref{fig_ref}(b). In case of the spherulite opal, however, there are regions of frequencies above $a/\lambda_{0} = 1.2$ which exhibit near-unity reflectivity as a consequence of the presence of pseudogaps. This occurs for both solid and core shell spherulite opals, as shown in Figs. \ref{fig_ref}(c) and \ref{fig_ref}(d), respectively. In addition to pseudogaps, it is predicted that the large group index associated with the nearly flat bands also leads to increased reflectivity \cite{sakoda2004optical}. The reflectivity at frequencies that lie in the lowest pseudogap is less than unity because the thickness of 18 layers of the opal at these wavelengths is not sufficient to fully reflect these frequencies. Simulating thicker structures accurately for these wavelengths using FDTD would cause the simulation times to be prohibitively large.

To summarize, we performed calculations of the photonic band structure and reflectivity of a spherulite opal. The results of our calculations indicate that the unique birefringent behaviour of spherulites results in the formation of new pseudogaps in the band structure, while preserving the mode symmetries of an opal made of optically isotropic spheres. For an interface parallel to the (111) plane, these pseudogaps result in enhanced reflectivity for frequencies above the third and fourth bands.  We calculated these effects using realistic material parameters inspired by recent findings in structural biology, and this provides motivation for studies of the optical properties of arrays of spherulites, and nanoscale patterned birefringent materials in general. Also, these results can inspire further exploration of biological systems for highly birefringent materials for applications in photonics.\\

Israeli Center of Research Excellence “Circle of Light” and the Crown Photonics Centre. D.O. is the incumbent of the Harry Weinrebe Professorial Chair of Laser Physics.

The authors thank Ron Tenne and Benjamin Palmer for helpful discussions.

\bigskip

\bibliography{refs}

\begin{thebibliography}{26}%
\makeatletter
\providecommand \@ifxundefined [1]{%
 \@ifx{#1\undefined}
}%
\providecommand \@ifnum [1]{%
 \ifnum #1\expandafter \@firstoftwo
 \else \expandafter \@secondoftwo
 \fi
}%
\providecommand \@ifx [1]{%
 \ifx #1\expandafter \@firstoftwo
 \else \expandafter \@secondoftwo
 \fi
}%
\providecommand \natexlab [1]{#1}%
\providecommand \enquote  [1]{``#1''}%
\providecommand \bibnamefont  [1]{#1}%
\providecommand \bibfnamefont [1]{#1}%
\providecommand \citenamefont [1]{#1}%
\providecommand \href@noop [0]{\@secondoftwo}%
\providecommand \href [0]{\begingroup \@sanitize@url \@href}%
\providecommand \@href[1]{\@@startlink{#1}\@@href}%
\providecommand \@@href[1]{\endgroup#1\@@endlink}%
\providecommand \@sanitize@url [0]{\catcode `\\12\catcode `\$12\catcode
  `\&12\catcode `\#12\catcode `\^12\catcode `\_12\catcode `\%12\relax}%
\providecommand \@@startlink[1]{}%
\providecommand \@@endlink[0]{}%
\providecommand \url  [0]{\begingroup\@sanitize@url \@url }%
\providecommand \@url [1]{\endgroup\@href {#1}{\urlprefix }}%
\providecommand \urlprefix  [0]{URL }%
\providecommand \Eprint [0]{\href }%
\providecommand \doibase [0]{http://dx.doi.org/}%
\providecommand \selectlanguage [0]{\@gobble}%
\providecommand \bibinfo  [0]{\@secondoftwo}%
\providecommand \bibfield  [0]{\@secondoftwo}%
\providecommand \translation [1]{[#1]}%
\providecommand \BibitemOpen [0]{}%
\providecommand \bibitemStop [0]{}%
\providecommand \bibitemNoStop [0]{.\EOS\space}%
\providecommand \EOS [0]{\spacefactor3000\relax}%
\providecommand \BibitemShut  [1]{\csname bibitem#1\endcsname}%
\let\auto@bib@innerbib\@empty
\bibitem [{\citenamefont {Yablonovitch}(1987)}]{yab1987}%
  \BibitemOpen
  \bibfield  {author} {\bibinfo {author} {\bibfnamefont {E.}~\bibnamefont
  {Yablonovitch}},\ }\href {\doibase 10.1103/PhysRevLett.58.2059} {\bibfield
  {journal} {\bibinfo  {journal} {Phys. Rev. Lett.}\ }\textbf {\bibinfo
  {volume} {58}},\ \bibinfo {pages} {2059} (\bibinfo {year}
  {1987})}\BibitemShut {NoStop}%
\bibitem [{\citenamefont {Joannopoulos}\ \emph {et~al.}(2011)\citenamefont
  {Joannopoulos}, \citenamefont {Johnson}, \citenamefont {Winn},\ and\
  \citenamefont {Meade}}]{book_joannopoulos}%
  \BibitemOpen
  \bibfield  {author} {\bibinfo {author} {\bibfnamefont {J.~D.}\ \bibnamefont
  {Joannopoulos}}, \bibinfo {author} {\bibfnamefont {S.~G.}\ \bibnamefont
  {Johnson}}, \bibinfo {author} {\bibfnamefont {J.~N.}\ \bibnamefont {Winn}}, \
  and\ \bibinfo {author} {\bibfnamefont {R.~D.}\ \bibnamefont {Meade}},\
  }\href@noop {} {\emph {\bibinfo {title} {Photonic Crystals: Molding the Flow
  of Light}}}\ (\bibinfo  {publisher} {Princeton University Press},\ \bibinfo
  {year} {2011})\BibitemShut {NoStop}%
\bibitem [{\citenamefont {Busch}\ and\ \citenamefont
  {John}(1998)}]{busch-john}%
  \BibitemOpen
  \bibfield  {author} {\bibinfo {author} {\bibfnamefont {K.}~\bibnamefont
  {Busch}}\ and\ \bibinfo {author} {\bibfnamefont {S.}~\bibnamefont {John}},\
  }\href {\doibase 10.1103/PhysRevE.58.3896} {\bibfield  {journal} {\bibinfo
  {journal} {Phys. Rev. E}\ }\textbf {\bibinfo {volume} {58}},\ \bibinfo
  {pages} {3896} (\bibinfo {year} {1998})}\BibitemShut {NoStop}%
\bibitem [{\citenamefont {Xia}\ \emph {et~al.}(2001)\citenamefont {Xia},
  \citenamefont {Gates},\ and\ \citenamefont {Li}}]{xia_review}%
  \BibitemOpen
  \bibfield  {author} {\bibinfo {author} {\bibfnamefont {Y.}~\bibnamefont
  {Xia}}, \bibinfo {author} {\bibfnamefont {B.}~\bibnamefont {Gates}}, \ and\
  \bibinfo {author} {\bibfnamefont {Z.-Y.}\ \bibnamefont {Li}},\ }\href
  {\doibase 10.1002/1521-4095(200103)13:6<409::AID-ADMA409>3.0.CO;2-C}
  {\bibfield  {journal} {\bibinfo  {journal} {Advanced Materials}\ }\textbf
  {\bibinfo {volume} {13}},\ \bibinfo {pages} {409} (\bibinfo {year}
  {2001})}\BibitemShut {NoStop}%
\bibitem [{\citenamefont {Kim}\ \emph {et~al.}(2011)\citenamefont {Kim},
  \citenamefont {Lee}, \citenamefont {Yang},\ and\ \citenamefont
  {Yi}}]{kim2011self}%
  \BibitemOpen
  \bibfield  {author} {\bibinfo {author} {\bibfnamefont {S.-H.}\ \bibnamefont
  {Kim}}, \bibinfo {author} {\bibfnamefont {S.~Y.}\ \bibnamefont {Lee}},
  \bibinfo {author} {\bibfnamefont {S.-M.}\ \bibnamefont {Yang}}, \ and\
  \bibinfo {author} {\bibfnamefont {G.-R.}\ \bibnamefont {Yi}},\ }\href@noop {}
  {\bibfield  {journal} {\bibinfo  {journal} {NPG Asia Materials}\ }\textbf
  {\bibinfo {volume} {3}},\ \bibinfo {pages} {25} (\bibinfo {year}
  {2011})}\BibitemShut {NoStop}%
\bibitem [{\citenamefont {Schroden}\ \emph {et~al.}(2002)\citenamefont
  {Schroden}, \citenamefont {Al-Daous}, \citenamefont {Blanford},\ and\
  \citenamefont {Stein}}]{stein_com}%
  \BibitemOpen
  \bibfield  {author} {\bibinfo {author} {\bibfnamefont {R.~C.}\ \bibnamefont
  {Schroden}}, \bibinfo {author} {\bibfnamefont {M.}~\bibnamefont {Al-Daous}},
  \bibinfo {author} {\bibfnamefont {C.~F.}\ \bibnamefont {Blanford}}, \ and\
  \bibinfo {author} {\bibfnamefont {A.}~\bibnamefont {Stein}},\ }\href
  {\doibase 10.1021/cm020100z} {\bibfield  {journal} {\bibinfo  {journal}
  {Chemistry of Materials}\ }\textbf {\bibinfo {volume} {14}},\ \bibinfo
  {pages} {3305} (\bibinfo {year} {2002})}\BibitemShut {NoStop}%
\bibitem [{\citenamefont {Aguirre}\ \emph {et~al.}(2010)\citenamefont
  {Aguirre}, \citenamefont {Reguera},\ and\ \citenamefont
  {Stein}}]{stein_adfm}%
  \BibitemOpen
  \bibfield  {author} {\bibinfo {author} {\bibfnamefont {C.~I.}\ \bibnamefont
  {Aguirre}}, \bibinfo {author} {\bibfnamefont {E.}~\bibnamefont {Reguera}}, \
  and\ \bibinfo {author} {\bibfnamefont {A.}~\bibnamefont {Stein}},\ }\href
  {\doibase 10.1002/adfm.201000143} {\bibfield  {journal} {\bibinfo  {journal}
  {Advanced Functional Materials}\ }\textbf {\bibinfo {volume} {20}},\ \bibinfo
  {pages} {2565} (\bibinfo {year} {2010})}\BibitemShut {NoStop}%
\bibitem [{\citenamefont {Li}\ \emph {et~al.}(1998{\natexlab{a}})\citenamefont
  {Li}, \citenamefont {Gu},\ and\ \citenamefont {Yang}}]{Li1998_2d}%
  \BibitemOpen
  \bibfield  {author} {\bibinfo {author} {\bibfnamefont {Z.-Y.}\ \bibnamefont
  {Li}}, \bibinfo {author} {\bibfnamefont {B.-Y.}\ \bibnamefont {Gu}}, \ and\
  \bibinfo {author} {\bibfnamefont {G.-Z.}\ \bibnamefont {Yang}},\ }\href@noop
  {} {\bibfield  {journal} {\bibinfo  {journal} {Physical Review Letters}\
  }\textbf {\bibinfo {volume} {81}},\ \bibinfo {pages} {2574} (\bibinfo {year}
  {1998}{\natexlab{a}})}\BibitemShut {NoStop}%
\bibitem [{\citenamefont {Li}\ \emph {et~al.}(1998{\natexlab{b}})\citenamefont
  {Li}, \citenamefont {Wang},\ and\ \citenamefont {Gu}}]{Li1998_3d}%
  \BibitemOpen
  \bibfield  {author} {\bibinfo {author} {\bibfnamefont {Z.-Y.}\ \bibnamefont
  {Li}}, \bibinfo {author} {\bibfnamefont {J.}~\bibnamefont {Wang}}, \ and\
  \bibinfo {author} {\bibfnamefont {B.-Y.}\ \bibnamefont {Gu}},\ }\href
  {\doibase 10.1103/PhysRevB.58.3721} {\bibfield  {journal} {\bibinfo
  {journal} {Phys. Rev. B}\ }\textbf {\bibinfo {volume} {58}},\ \bibinfo
  {pages} {3721} (\bibinfo {year} {1998}{\natexlab{b}})}\BibitemShut {NoStop}%
\bibitem [{\citenamefont {Zabel}\ and\ \citenamefont
  {Stroud}(1993)}]{zabel_stroud}%
  \BibitemOpen
  \bibfield  {author} {\bibinfo {author} {\bibfnamefont {I.~H.~H.}\
  \bibnamefont {Zabel}}\ and\ \bibinfo {author} {\bibfnamefont
  {D.}~\bibnamefont {Stroud}},\ }\href {\doibase 10.1103/PhysRevB.48.5004}
  {\bibfield  {journal} {\bibinfo  {journal} {Phys. Rev. B}\ }\textbf {\bibinfo
  {volume} {48}},\ \bibinfo {pages} {5004} (\bibinfo {year}
  {1993})}\BibitemShut {NoStop}%
\bibitem [{\citenamefont {Busch}\ and\ \citenamefont
  {John}(1999)}]{busch-john-lc}%
  \BibitemOpen
  \bibfield  {author} {\bibinfo {author} {\bibfnamefont {K.}~\bibnamefont
  {Busch}}\ and\ \bibinfo {author} {\bibfnamefont {S.}~\bibnamefont {John}},\
  }\href {\doibase 10.1103/PhysRevLett.83.967} {\bibfield  {journal} {\bibinfo
  {journal} {Phys. Rev. Lett.}\ }\textbf {\bibinfo {volume} {83}},\ \bibinfo
  {pages} {967} (\bibinfo {year} {1999})}\BibitemShut {NoStop}%
\bibitem [{\citenamefont {Palmer}\ \emph {et~al.}(2018)\citenamefont {Palmer},
  \citenamefont {Hirsch}, \citenamefont {Brumfeld}, \citenamefont {Aflalo},
  \citenamefont {Pinkas}, \citenamefont {Sagi}, \citenamefont {Rosenne},
  \citenamefont {Oron}, \citenamefont {Leiserowitz}, \citenamefont {Kronik},
  \citenamefont {Weiner},\ and\ \citenamefont {Addadi}}]{ben_pnas}%
  \BibitemOpen
  \bibfield  {author} {\bibinfo {author} {\bibfnamefont {B.~A.}\ \bibnamefont
  {Palmer}}, \bibinfo {author} {\bibfnamefont {A.}~\bibnamefont {Hirsch}},
  \bibinfo {author} {\bibfnamefont {V.}~\bibnamefont {Brumfeld}}, \bibinfo
  {author} {\bibfnamefont {E.~D.}\ \bibnamefont {Aflalo}}, \bibinfo {author}
  {\bibfnamefont {I.}~\bibnamefont {Pinkas}}, \bibinfo {author} {\bibfnamefont
  {A.}~\bibnamefont {Sagi}}, \bibinfo {author} {\bibfnamefont {S.}~\bibnamefont
  {Rosenne}}, \bibinfo {author} {\bibfnamefont {D.}~\bibnamefont {Oron}},
  \bibinfo {author} {\bibfnamefont {L.}~\bibnamefont {Leiserowitz}}, \bibinfo
  {author} {\bibfnamefont {L.}~\bibnamefont {Kronik}}, \bibinfo {author}
  {\bibfnamefont {S.}~\bibnamefont {Weiner}}, \ and\ \bibinfo {author}
  {\bibfnamefont {L.}~\bibnamefont {Addadi}},\ }\href {\doibase
  10.1073/pnas.1722531115} {\bibfield  {journal} {\bibinfo  {journal}
  {Proceedings of the National Academy of Sciences of the United States of
  America}\ }\textbf {\bibinfo {volume} {115}},\ \bibinfo {pages} {2299}
  (\bibinfo {year} {2018})}\BibitemShut {NoStop}%
\bibitem [{\citenamefont {Palmer}\ \emph {et~al.}(2019)\citenamefont {Palmer},
  \citenamefont {Yallapragada}, \citenamefont {Schiffmann}, \citenamefont
  {Wormser}, \citenamefont {Elad}, \citenamefont {Aflalo}, \citenamefont
  {Sagi}, \citenamefont {Weiner}, \citenamefont {Addadi},\ and\ \citenamefont
  {Oron}}]{decapod_paper}%
  \BibitemOpen
  \bibfield  {author} {\bibinfo {author} {\bibfnamefont {B.~A.}\ \bibnamefont
  {Palmer}}, \bibinfo {author} {\bibfnamefont {V.~J.}\ \bibnamefont
  {Yallapragada}}, \bibinfo {author} {\bibfnamefont {N.}~\bibnamefont
  {Schiffmann}}, \bibinfo {author} {\bibfnamefont {E.~M.}\ \bibnamefont
  {Wormser}}, \bibinfo {author} {\bibfnamefont {N.}~\bibnamefont {Elad}},
  \bibinfo {author} {\bibfnamefont {E.~D.}\ \bibnamefont {Aflalo}}, \bibinfo
  {author} {\bibfnamefont {A.}~\bibnamefont {Sagi}}, \bibinfo {author}
  {\bibfnamefont {S.}~\bibnamefont {Weiner}}, \bibinfo {author} {\bibfnamefont
  {L.}~\bibnamefont {Addadi}}, \ and\ \bibinfo {author} {\bibfnamefont
  {D.}~\bibnamefont {Oron}},\ }\href@noop {} {\bibfield  {journal} {\bibinfo
  {journal} {Submitted}\ } (\bibinfo {year} {2019})}\BibitemShut {NoStop}%
\bibitem [{\citenamefont {Bower}(2002)}]{bower_book}%
  \BibitemOpen
  \bibfield  {author} {\bibinfo {author} {\bibfnamefont {D.~I.}\ \bibnamefont
  {Bower}},\ }\href@noop {} {\emph {\bibinfo {title} {An Introduction to
  Polymer Physics}}}\ (\bibinfo  {publisher} {Cambridge University Press},\
  \bibinfo {year} {2002})\BibitemShut {NoStop}%
\bibitem [{\citenamefont {Roth}\ and\ \citenamefont
  {Dignam}(1973)}]{roth_dignam}%
  \BibitemOpen
  \bibfield  {author} {\bibinfo {author} {\bibfnamefont {J.}~\bibnamefont
  {Roth}}\ and\ \bibinfo {author} {\bibfnamefont {M.~J.}\ \bibnamefont
  {Dignam}},\ }\href {\doibase 10.1364/JOSA.63.000308} {\bibfield  {journal}
  {\bibinfo  {journal} {J. Opt. Soc. Am.}\ }\textbf {\bibinfo {volume} {63}},\
  \bibinfo {pages} {308} (\bibinfo {year} {1973})}\BibitemShut {NoStop}%
\bibitem [{\citenamefont {Ni}\ \emph {et~al.}(2013)\citenamefont {Ni},
  \citenamefont {Gao}, \citenamefont {Miroshnichenko},\ and\ \citenamefont
  {Qiu}}]{ni_spherulite}%
  \BibitemOpen
  \bibfield  {author} {\bibinfo {author} {\bibfnamefont {Y.~X.}\ \bibnamefont
  {Ni}}, \bibinfo {author} {\bibfnamefont {L.}~\bibnamefont {Gao}}, \bibinfo
  {author} {\bibfnamefont {A.~E.}\ \bibnamefont {Miroshnichenko}}, \ and\
  \bibinfo {author} {\bibfnamefont {C.~W.}\ \bibnamefont {Qiu}},\ }\href
  {\doibase 10.1364/OE.21.008091} {\bibfield  {journal} {\bibinfo  {journal}
  {Opt. Express}\ }\textbf {\bibinfo {volume} {21}},\ \bibinfo {pages} {8091}
  (\bibinfo {year} {2013})}\BibitemShut {NoStop}%
\bibitem [{\citenamefont {Kiselev}\ \emph {et~al.}(2002)\citenamefont
  {Kiselev}, \citenamefont {Reshetnyak},\ and\ \citenamefont
  {Sluckin}}]{kiselev_tmatrix}%
  \BibitemOpen
  \bibfield  {author} {\bibinfo {author} {\bibfnamefont {A.~D.}\ \bibnamefont
  {Kiselev}}, \bibinfo {author} {\bibfnamefont {V.~Y.}\ \bibnamefont
  {Reshetnyak}}, \ and\ \bibinfo {author} {\bibfnamefont {T.~J.}\ \bibnamefont
  {Sluckin}},\ }\href {\doibase 10.1103/PhysRevE.65.056609} {\bibfield
  {journal} {\bibinfo  {journal} {Phys. Rev. E}\ }\textbf {\bibinfo {volume}
  {65}},\ \bibinfo {pages} {056609} (\bibinfo {year} {2002})}\BibitemShut
  {NoStop}%
\bibitem [{\citenamefont {Johnson}\ and\ \citenamefont
  {Joannopoulos}(2001)}]{mpb}%
  \BibitemOpen
  \bibfield  {author} {\bibinfo {author} {\bibfnamefont {S.~G.}\ \bibnamefont
  {Johnson}}\ and\ \bibinfo {author} {\bibfnamefont {J.~D.}\ \bibnamefont
  {Joannopoulos}},\ }\href
  {http://www.opticsexpress.org/abstract.cfm?URI=OPEX-8-3-173} {\bibfield
  {journal} {\bibinfo  {journal} {Opt. Express}\ }\textbf {\bibinfo {volume}
  {8}},\ \bibinfo {pages} {173} (\bibinfo {year} {2001})}\BibitemShut {NoStop}%
\bibitem [{\citenamefont {Bogomolov}\ \emph {et~al.}(1997)\citenamefont
  {Bogomolov}, \citenamefont {Gaponenko}, \citenamefont {Germanenko},
  \citenamefont {Kapitonov}, \citenamefont {Petrov}, \citenamefont {Gaponenko},
  \citenamefont {Prokofiev}, \citenamefont {Ponyavina}, \citenamefont
  {Silvanovich},\ and\ \citenamefont {Samoilovich}}]{bogomolov}%
  \BibitemOpen
  \bibfield  {author} {\bibinfo {author} {\bibfnamefont {V.~N.}\ \bibnamefont
  {Bogomolov}}, \bibinfo {author} {\bibfnamefont {S.~V.}\ \bibnamefont
  {Gaponenko}}, \bibinfo {author} {\bibfnamefont {I.~N.}\ \bibnamefont
  {Germanenko}}, \bibinfo {author} {\bibfnamefont {A.~M.}\ \bibnamefont
  {Kapitonov}}, \bibinfo {author} {\bibfnamefont {E.~P.}\ \bibnamefont
  {Petrov}}, \bibinfo {author} {\bibfnamefont {N.~V.}\ \bibnamefont
  {Gaponenko}}, \bibinfo {author} {\bibfnamefont {A.~V.}\ \bibnamefont
  {Prokofiev}}, \bibinfo {author} {\bibfnamefont {A.~N.}\ \bibnamefont
  {Ponyavina}}, \bibinfo {author} {\bibfnamefont {N.~I.}\ \bibnamefont
  {Silvanovich}}, \ and\ \bibinfo {author} {\bibfnamefont {S.~M.}\ \bibnamefont
  {Samoilovich}},\ }\href {\doibase 10.1103/PhysRevE.55.7619} {\bibfield
  {journal} {\bibinfo  {journal} {Phys. Rev. E}\ }\textbf {\bibinfo {volume}
  {55}},\ \bibinfo {pages} {7619} (\bibinfo {year} {1997})}\BibitemShut
  {NoStop}%
\bibitem [{\citenamefont {Pavarini}\ \emph {et~al.}(2005)\citenamefont
  {Pavarini}, \citenamefont {Andreani}, \citenamefont {Soci}, \citenamefont
  {Galli}, \citenamefont {Marabelli},\ and\ \citenamefont
  {Comoretto}}]{pavarini}%
  \BibitemOpen
  \bibfield  {author} {\bibinfo {author} {\bibfnamefont {E.}~\bibnamefont
  {Pavarini}}, \bibinfo {author} {\bibfnamefont {L.~C.}\ \bibnamefont
  {Andreani}}, \bibinfo {author} {\bibfnamefont {C.}~\bibnamefont {Soci}},
  \bibinfo {author} {\bibfnamefont {M.}~\bibnamefont {Galli}}, \bibinfo
  {author} {\bibfnamefont {F.}~\bibnamefont {Marabelli}}, \ and\ \bibinfo
  {author} {\bibfnamefont {D.}~\bibnamefont {Comoretto}},\ }\href {\doibase
  10.1103/PhysRevB.72.045102} {\bibfield  {journal} {\bibinfo  {journal} {Phys.
  Rev. B}\ }\textbf {\bibinfo {volume} {72}},\ \bibinfo {pages} {045102}
  (\bibinfo {year} {2005})}\BibitemShut {NoStop}%
\bibitem [{\citenamefont {Sakoda}(2004)}]{sakoda2004optical}%
  \BibitemOpen
  \bibfield  {author} {\bibinfo {author} {\bibfnamefont {K.}~\bibnamefont
  {Sakoda}},\ }\href@noop {} {\emph {\bibinfo {title} {Optical properties of
  photonic crystals}}}\ (\bibinfo  {publisher} {Springer},\ \bibinfo {year}
  {2004})\BibitemShut {NoStop}%
\bibitem [{\citenamefont {L\'opez-Tejeira}\ \emph {et~al.}(2002)\citenamefont
  {L\'opez-Tejeira}, \citenamefont {Ochiai}, \citenamefont {Sakoda},\ and\
  \citenamefont {S\'anchez-Dehesa}}]{sakoda_prb}%
  \BibitemOpen
  \bibfield  {author} {\bibinfo {author} {\bibfnamefont {F.}~\bibnamefont
  {L\'opez-Tejeira}}, \bibinfo {author} {\bibfnamefont {T.}~\bibnamefont
  {Ochiai}}, \bibinfo {author} {\bibfnamefont {K.}~\bibnamefont {Sakoda}}, \
  and\ \bibinfo {author} {\bibfnamefont {J.}~\bibnamefont {S\'anchez-Dehesa}},\
  }\href {\doibase 10.1103/PhysRevB.65.195110} {\bibfield  {journal} {\bibinfo
  {journal} {Phys. Rev. B}\ }\textbf {\bibinfo {volume} {65}},\ \bibinfo
  {pages} {195110} (\bibinfo {year} {2002})}\BibitemShut {NoStop}%
\bibitem [{\citenamefont {Taflove}\ and\ \citenamefont
  {Hagness}(2005)}]{taflove}%
  \BibitemOpen
  \bibfield  {author} {\bibinfo {author} {\bibfnamefont {A.}~\bibnamefont
  {Taflove}}\ and\ \bibinfo {author} {\bibfnamefont {S.~C.}\ \bibnamefont
  {Hagness}},\ }\href@noop {} {\emph {\bibinfo {title} {Computational
  electrodynamics: the finite-difference time-domain method}}}\ (\bibinfo
  {publisher} {Artech house},\ \bibinfo {year} {2005})\BibitemShut {NoStop}%
\bibitem [{lum()}]{lumerical}%
  \BibitemOpen
  \href@noop {} {\enquote {\bibinfo {title} {Lumerical inc.}}\ }\bibinfo
  {howpublished} {{https://www.lumerical.com/products/fdtd/}}\BibitemShut
  {NoStop}%
\bibitem [{\citenamefont {Balestreri}\ \emph {et~al.}(2006)\citenamefont
  {Balestreri}, \citenamefont {Andreani},\ and\ \citenamefont {Agio}}]{agio}%
  \BibitemOpen
  \bibfield  {author} {\bibinfo {author} {\bibfnamefont {A.}~\bibnamefont
  {Balestreri}}, \bibinfo {author} {\bibfnamefont {L.~C.}\ \bibnamefont
  {Andreani}}, \ and\ \bibinfo {author} {\bibfnamefont {M.}~\bibnamefont
  {Agio}},\ }\href {\doibase 10.1103/PhysRevE.74.036603} {\bibfield  {journal}
  {\bibinfo  {journal} {Phys. Rev. E}\ }\textbf {\bibinfo {volume} {74}},\
  \bibinfo {pages} {036603} (\bibinfo {year} {2006})}\BibitemShut {NoStop}%
\bibitem [{\citenamefont {Galisteo-L\'opez}\ and\ \citenamefont
  {L\'opez}(2004)}]{galisteo_high_energy}%
  \BibitemOpen
  \bibfield  {author} {\bibinfo {author} {\bibfnamefont {J.~F.}\ \bibnamefont
  {Galisteo-L\'opez}}\ and\ \bibinfo {author} {\bibfnamefont {C.}~\bibnamefont
  {L\'opez}},\ }\href {\doibase 10.1103/PhysRevB.70.035108} {\bibfield
  {journal} {\bibinfo  {journal} {Phys. Rev. B}\ }\textbf {\bibinfo {volume}
  {70}},\ \bibinfo {pages} {035108} (\bibinfo {year} {2004})}\BibitemShut
  {NoStop}%
\end{thebibliography}%
\end{document}